\documentclass[a4paper,fleqn,usenatbib]{mnras}

\usepackage{graphicx}	
\usepackage{amsmath}	
\usepackage{amssymb}	
\usepackage{hyperref}
\usepackage{bm}
\title[Human and machine classifications]{A transient search using combined human and machine classifications}

\author[D.E. Wright, C. J. Lintott, S.J. Smartt  et al.]{Darryl E. Wright,$^{1}$\thanks{E-mail: darryl@zooniverse.org}
Chris J. Lintott,$^{1}$
Stephen J. Smartt,$^{2}$
Ken W. Smith,$^{2}$\newauthor
Lucy Fortson,$^{3,4}$
Laura Trouille,$^{5,6}$
Campbell R. Allen,$^{1}$
Melanie Beck,$^{3}$\newauthor
Mark C. Bouslog,$^{6}$
Amy Boyer,$^{6}$
K. C. Chambers,$^{7}$
Heather Flewelling,$^{7}$\newauthor
Will Granger,$^{6}$
Eugene A. Magnier,$^{7}$
Adam McMaster,$^{1}$
Grant R. M. Miller,$^{1}$\newauthor
James E. O'Donnell,$^{1}$
Helen Spiers,$^{1}$
John L. Tonry,$^{7}$
Marten Veldthuis,$^{1}$\newauthor
Richard J. Wainscoat,$^{7}$
Chris Waters,$^{7}$
Mark Willman,$^{7}$
Zach Wolfenbarger,${^6}$\newauthor
Dave R. Young$^{2}$\\
\\
$^{1}$Department of Physics, University of Oxford, Denys Wilkinson Building, Keble Road, Oxford, OX1 3RH, UK \\
$^{2}$Astrophysics Research Centre, School of Mathematics and Physics, Queen'€™s University Belfast, Belfast BT7 1NN, UK \\
$^{3}$Minnesota Institute for Astrophysics, University of Minnesota, Minneapolis, MN 55454, USA \\
$^{4}$School of Physics and Astronomy, University of Minnesota, Minneapolis, MN 55455, USA\\
$^{5}$Center for Interdisciplinary Exploration and Research in Astrophysics (CIERA) and Dept. of Physics and Astronomy, \\
\hspace{2pt} Northwestern University, Evanston, IL, USA\\
$^{6}$Citizen Science Department, The Adler Planetarium, Chicago, IL, USA\\
$^{7}$Institute for Astronomy, University of Hawaii, 2680 Woodlawn Drive, Honolulu, HI 96822, USA \\
}

\date{Accepted XXX. Received YYY; in original form ZZZ}

\pubyear{2016}

\begin{document}
\label{firstpage}
\pagerange{\pageref{firstpage}--\pageref{lastpage}}
\maketitle

\begin{abstract}
Large modern surveys require efficient review of data in order to find transient sources such as supernovae, and to distinguish such sources from artefacts and noise. Much effort has been put into the development of automatic algorithms, but surveys still rely on human review of targets. This paper presents an integrated system for the identification of supernovae in data from Pan-STARRS1, combining classifications from volunteers participating in a citizen science project with those from a convolutional neural network. The unique aspect of this work is the deployment, in combination, of both human and machine classifications for near real-time discovery in an astronomical project.  We show that the combination of the two methods outperforms either one used individually. This result has important implications for the future development of transient searches, especially in the era of LSST and other large-throughput surveys. 
\end{abstract}




\section{Introduction}

The detection and identification of transient sources has long been an important part of astronomical observation. New surveys such as LSST (Large Synoptic Survey Telescope, \citet{Ivezic08}) will increase the number of transient candidates detected by many orders of magnitude, leading to renewed attention being paid to the methods used by transient searches.  To extract the most scientific value from surveys, we want to follow the entire evolution of transients from the time of outburst to the point at which they fade below
 the detection limit.  This requires a rapid processing of data to enable a fast decision on whether or not to expend valuable follow up resources for each potential candidate extracted by a transient survey's image processing pipeline.  The first problem is deciding if a source, flagged by the pipeline, is a detection with real astrophysical significance or an artefact of the detector or image processing.  We want to promote the former for a decision on whether to followup  and to reject the latter without further consideration.  In preparation for LSST and to deal with the data volumes of present surveys, much effort has been invested in developing systems which automatically reject false positives with supervised learning. Using large volumes of past observations that have been identified as real or ``bogus'', the aim is to train a machine to make predictions about future observations \citep{Bloom12, Brink13, Goldstein15, duBuisson15, Donalek08, Romano06, Bailey07, Masci17}.  Providing this data typically requires manual inspection of individual detections by human experts to mitigate label contamination that would confound the learning algorithm.  This quickly becomes unwieldy given that performance of a machine learning solution has been shown to depend on the quantity of labelled training data \citep{Banko01}.

The requirement for sufficiently large and representative training sets is prohibitive for the largest surveys or for small research teams, and is particularly problematic for rarer classes of transients.  An alternative to relying on expert labelled training sets for machine learning is therefore to expand the population providing labels.
 For example \citet{Melchior16} describes a crowd sourcing platform for vetting image quality from the Dark Energy Survey (DES) \citep{dark2005} relying on the consensus of $\sim100$ volunteers from a team of professional astronomers.  For surveys of the future a few hundred volunteers will not be enough.  Instead we must cast our net wider increasing the population of the crowd beyond those directly involved in the survey.  The obvious path is to engage citizen scientists.  \textit{Galaxy Zoo Supernovae} \citep{Smith11} and \textit{Snapshot Supernova} \citep{Campbell15} are two projects to have taken this approach for transient surveys, using data from the Palomar Transient Factory (PTF) \citep{Rau09, Law09} and SkyMapper \citep{Keller07} respectively.  Both projects were facilitated through the \textit{Zooniverse} Citizen Science platform (see description in \citet{Marshall15}), and asked volunteers to assess the target, reference and difference images for each detection and answer a series of questions that led to a classification of real or bogus.  

For classification tasks, humans and machines have complementary strengths.  Human classifiers are good at rapidly making abstract judgments about data, allowing them to see only a small number of examples before making decisions about novel images.  Machines can consume large quantities of data and make more systematic judgements based on complex relationships between the features provided.  In the \textit{Space Warps} citizen science project \citep{Marshall16} all example gravitational lenses provided to volunteers appeared blue, yet despite this, volunteers were able to identify a gravitationally lensed hyperluminous infrared radio galaxy that appeared red \citep{Geach15}.  In contrast a machine would need to be provided examples of these in the training data. On the other hand, computer vision techniques allow images to be examined systematically, with relationships between different features used for classification. Combining machine classifications with those of experts will trivially be expected to improve performance, but the situation in which classifications from volunteer citizen scientists are used is less clear. If the noisier data sets provided by citizen science are combined with machines, is performance of the system improved? If machines are included in classification, does it relieve some of the burden on citizen scientists? Answering these questions is critically important for surveys where even a substantial number of experts will not be able to review all the data promoted by a machine classifier. 

In this paper we report some initial findings from the \textit{Supernova Hunters} project\footnote{\url{https://www.zooniverse.org/projects/dwright04/supernova-hunters}}, a new citizen science project similar in spirit to those mentioned above but applied to the Pan-STARRS Survey for Transients (PSST).  In Section~\ref{sec:method} we describe the Pan-STARRS1 telescope, PSST survey and the Supernova Hunters project and citizen science platform.  Section~\ref{sec:perform} shows the relative performance of humans and machines on data uploaded to Supernova Hunters during the first two months of the project.  We also describe and measure the performance of a simple method for combining the classifications of citizen scientists and the current PSST machine classifier.  We further discuss a mechanism to take advantage of metadata associated with each detection to boost classification performance.  In Section~\ref{sec:conclusions} we conclude and discuss potential avenues for future improvements. This paper therefore represents the first study of combined citizen science and machine classifications within a live astronomical survey. 

\section{Method}
\label{sec:method}
\subsection{Pan-STARRS1}
\label{sec:ps1}

Pan-STARRS1 comprises a 1.8m primary mirror \citep{Kaiser10} and 60 detectors with $4800\time 4800$ pixels, constructed from $10\mu m$ pixels subtending 0.258 arcsec \citep{Magnier13} and a field-of-view of 3.3 deg.  The filter set consists of $g_{P1}$, $r_{P1}$, $i_{P1}$, $z_{P1}$ (similar to SDSS \textit{griz} \citep{York00}), $y_{P1}$ extending
redward of $z_{P1}$ and the ``wide'' $w_{P1}$-band filter extending over $g_{P1}$ to $i_{P1}$ \citep{Tonry12b}.  Between 2010 and 2014 Pan-STARRS1 was operated by the PS1 Science Consortium (PS1SC) performing 2 major surveys.  The Medium Deep Survey (MDS) was allocated 25\% of observing time for high cadence observations of the 10 Medium Deep fields and the $3\pi$ survey allocated 56\% observing time to observe the entire sky north of -30 degrees declination with 4 exposures per year in each of $g_{P1}$, $r_{P1}$, $i_{P1}$, $z_{P1}$ and $y_{P1}$ for each pointing.

The $3\pi$ survey was completed in mid-2014 and since then the telescope has been carrying out a NASA funded wide-field survey for near earth objects through the NEO Observation Program operated by the Pan-STARRS Near Earth Object Science Consortium (PSNSC).  The NASA PSNSC survey is similar to the $3\pi$ survey but optimised for NEO discoveries.  Observations are in $w_{P1}$ in dark time and combinations of $i_{P1}$, $z_{P1}$ and $y_{P1}$ during bright time.  The PanSTARRS Survey for Transients (PSST) \citep{Huber15a, Inserra13} searches the data for static transients, releasing these publicly within 12 to 24 hours.

Typically a single field is imaged 4 times in a night with exposures separated by 10-20 mins called Transient Time Interval (TTI) exposures to allow for the discovery of moving objects.  The quads of exposures are not dithered or stacked, meaning that cross-talk ghosts, readout artefacts and problems of fill-factor are inherent in the data (see \citet{Denneau13} for some examples).  Individual exposures are differenced \citep{Alard98, Bramich08} with the $3\pi$ all-sky reference stack and sources in the resulting difference images are catalogued.

A series of pre-ingest cuts are performed before the catalogues are ingested into a MySQL database at Queen's University Belfast (QUB).  These cuts are based on the detection of saturated, masked or suspected defective pixels within the PSF area in addition to flag checks for ghost detections and rejecting detections within $\pm 5$ degrees galactic latitude. Detections passing these cuts are grouped into transient candidates if they are spatially coincident within 0.5 arcsec and the rms scatter is $<$ 0.25 arcsec.  Post-ingest cuts are applied on detection quality, convolution checks and a check for proximity to bright objects.  Additional cross-talk rules have been identified and implemented at QUB to reject ghosts
not flagged at the pre-ingest stage.  Remaining detections are cross-matched with the Minor Planets Center ephemeris database to identify any asteroids not 
removed by the rms cut.  Remaining transient candidates are passed to our machine classifier described in the next section.

\subsection{Convolutional Neural Network}
\label{sec:cnn}

In \citet{Wright15} we developed a machine classifier for real-bogus classification in the PS1 Medium Deep Survey.  However, we found that this approach performs poorly for PSST because of the greater variety of artefacts in PSST data (a consequence of differencing individual exposures) and the difficulty obtaining a representative labelled training set at the beginning of a new survey.  Instead we turned to Convolutional Neural Networks (CNNs) that maintain the advantages of operating solely on the pixel data but at a higher computational cost in deployment.  

The training set for this classifier was drawn from $3\pi$ survey data between 1st June 2013 and 20th June 2014.  The sample of real detections are taken from spectroscopically confirmed transients or detections of objects that have been labelled by experts as high probability real transients.  Bogus detections are taken from a random subsample of detections discarded by post-ingest cuts or human inspection.  The training set consists of 6916 examples with an additional 2303 detections held out for testing with both data sets containing twice as many bogus detections to real. Each example was manually inspected in order to limit label contamination; not all detections associated with a spectroscopically confirmed transient are necessarily real for example.

Given the small data set, to avoid overfitting we limit the CNN to a single convolution layer with 400 kernels and a pooling layer followed by a binary softmax classifier.  We also perform unsupervised pre-training with sparse filtering \citep{Ngiam11} using unlabelled images from the STL-10 \citep{Coates11} data set.  The classifier is applied to nightly PSST data producing a score for each TTI exposure for every candidate passing the cuts in the previous section.  The score, or \textit{hypothesis}, is a function $h(x)$ of the input feature representation, $x$ (the output of the convolution and pooling layers).  For each candidate we simply combine the TTI exposure hypotheses by taking the median, resulting in a single ``real-bogus factor'' for each transient candidate which we take as the machine equivalent of $P(real)$ below.  To automatically reject candidates we must choose a decision boundary on $h(x)$ such that any candidate with a hypothesis lying below the decision boundary is considered a bogus detection and discarded.  This inevitably leads to a trade-off between false positives and false negatives (or missed detections).  If the decision boundary is set too high we will discard many real detections of supernovae; too low and we will be inundated with artefacts (see for example Figure~\ref{fig:machine_dist}).  We chose the decision boundary based on the expected performance measured on the test set.  For example, using our CNN to generate hypotheses for each detection in the test set, we can choose the decision boundary that corresponds to a False Positive Rate (FPR) of 1\% at $h(x)=0.842$, where the FPR is the number of false positives divided by the total number of bogus candidates in the sample.  However, although the number of artefacts promoted would be low, based on the test set this decision boundary would be expected to result in a Missed Detection Rate (MDR) of $\sim21\%$ for future data and is therefore not a sensible choice.  We instead opted for a decision boundary at $h(x)=0.436$ with expected FPR and MDR of 5\% and $\sim$5.2\% respectively.  As detailed in Section~\ref{sec:combo} clearly the decision boundary can be scaled to take advantage of available human effort; lowering the decision boundary beyond 0.436 would increase the FPR requiring more human screening but at the same time reduce the MDR such that humans could recover real supernova detections with low $h(x)$ that would otherwise be automatically rejected.

\subsection{Citizen Science Platform}

\begin{figure*}
   \begin{minipage}{140mm}
   \includegraphics[width=140mm]{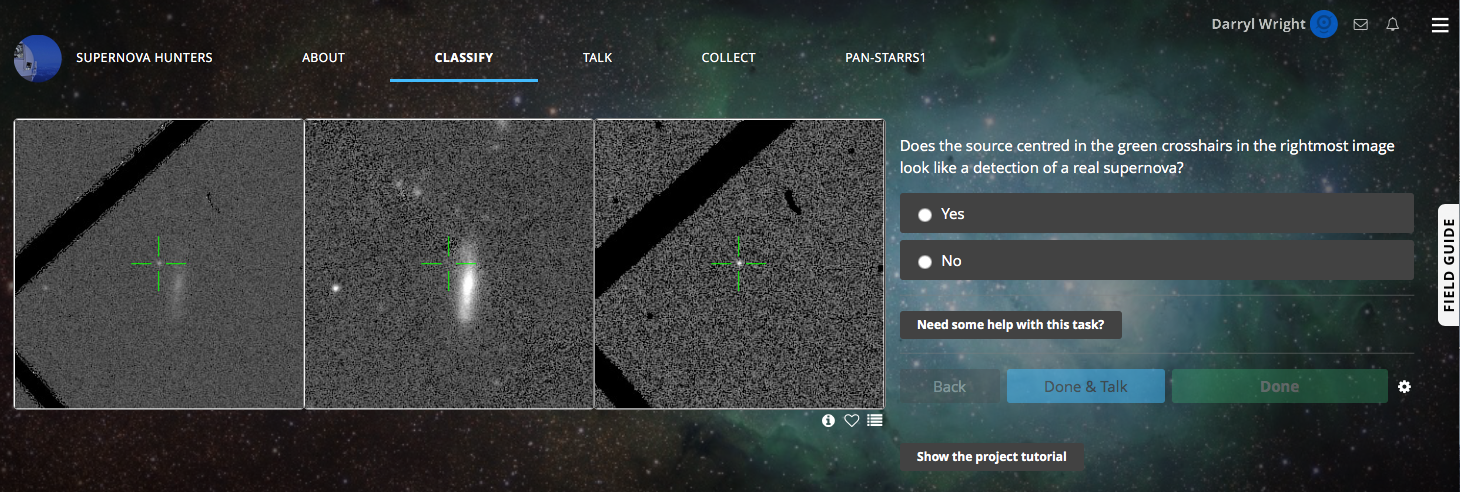}
   \caption{The classification interface presented to citizen scientists.  The left most image is the \textit{target} image taken during the previous week.  In the centre is the equivalent $3\pi$ \textit{reference} image and on the right is the \textit{difference} image.  Volunteers are asked to decide whether or not they think the detection in the green crosshairs in the difference image is a
detection of a real transient.}
   \label{fig:sn_hunters}
   \end{minipage}
\end{figure*}

Supernova Hunters was launched on 12th July 2016 (MJD 57581).  As of 6th December 2016 the project has accumulated 1082170 classifications from 5845 citizen scientists with a few tens of volunteers submitting thousands of classifications.  Citizen scientists are presented with the interface shown in Figure~\ref{fig:sn_hunters} and asked to classify individual TTI observations (see Section~\ref{sec:ps1}).  So far volunteers have classified 117693 individual images of 46277 individual PS1 objects. As guidance we provide a ``Field Guide'' that provides a description and examples of the different artefact types we expect.  Once a week $\sim5800$ new subjects are uploaded to the project consisting of the previous week's detections that pass our machine cuts. The arrival of the data is announced to existing volunteers via email\footnote{Though enthusiasm for the project is such that traffic to the site now increases significantly \emph{ahead} of this email alert!}. We require at least seven citizen scientist classifications before a subject is considered classified and subsequently retired from the project.  The choice of seven classifications is simply motivated by a trade-off between speed of data processing and accuracy and we did not see significant gains by requiring ten classifications during the beta test.  Since launch the project averages $\sim21000$ classifications in the first 24 hours after the data is released and $\sim8200$ classifications in the following 24 hours by which time all subjects are normally retired.

We calculate a ``human score'' simply by taking the fraction of all the volunteers who saw a detection and classified it was real.  High confidence (typically $P(real)>0.8$, see Section~\ref{sec:perform}) supernova candidates are screened by experts to remove a small number of false positives ($\sim10\%$) before the targets are submitted to the Transient Name Server (TNS).  To date citizen scientists have discovered over 450 supernova candidates that have been submitted to the TNS and two confirmed Supernovae including SN 2016els \citep{Mattila16}; a superluminous supernova Type I.  The classification spectra were obtained by PESSTO \citep{Smartt15}.

\section{Performance}
\label{sec:perform}

\begin{figure}
   \includegraphics[width=84mm]{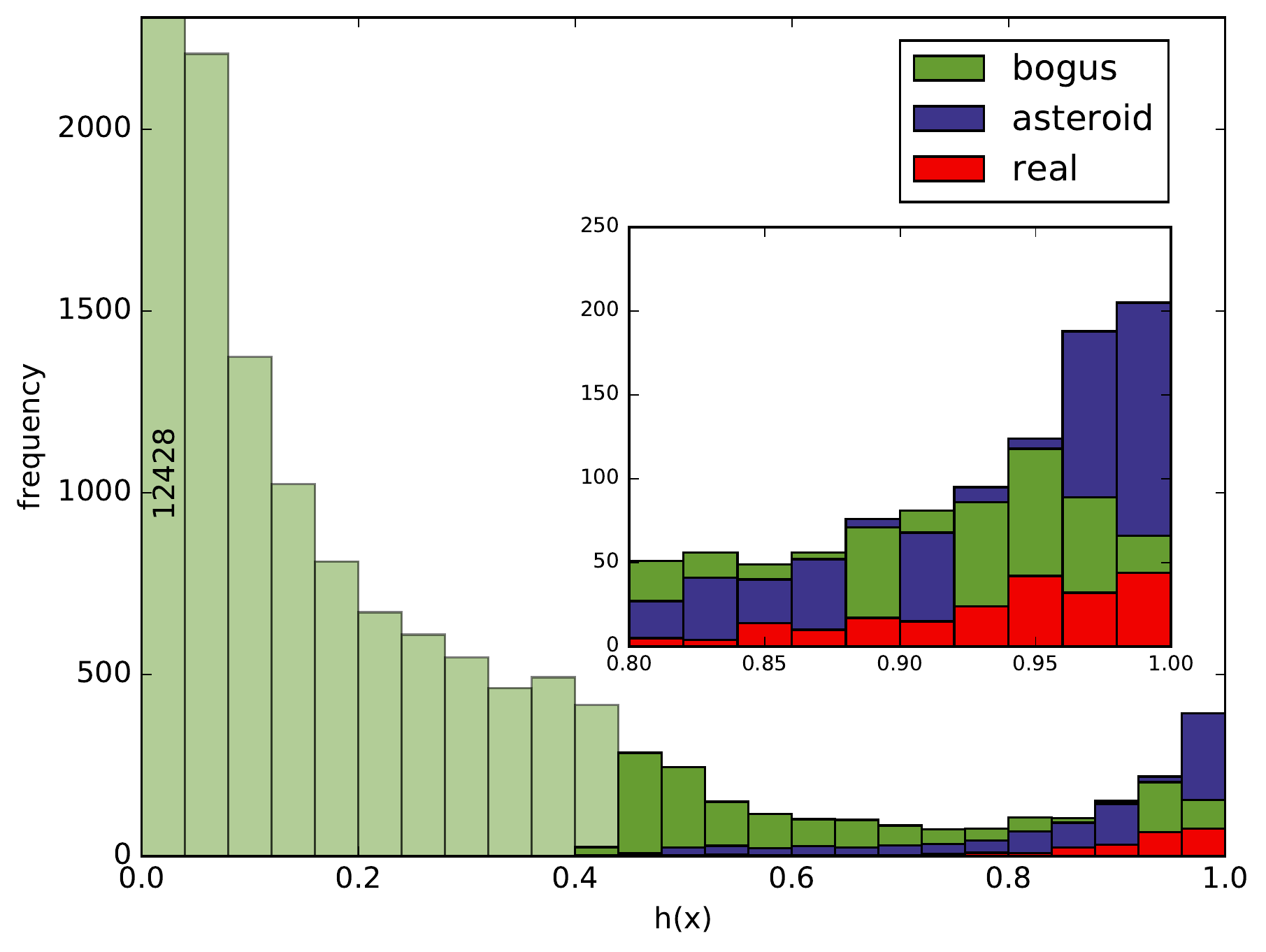}
   \caption{The distribution of hypotheses, $h(x)$ from the current 3$\pi$ machine classifier 
            for detected objects between MJD 57570 and MJD 57586.  The light green shows the distribution of 
            objects with $h(x) \leq 0.436$ which are automatically rejected.  The remaining 
            objects promoted for human screening even at high values of $h(x)$ contains
            many false positives.  The first interval has a frequency of 12428, but the plot is truncated for clarity. Inset: Zoom-in of the region with $h(x)>0.8$.} 
   \label{fig:machine_dist} 
\end{figure}

We present the results of our machine classifier (Section~\ref{sec:cnn}) in Figure~\ref{fig:machine_dist} on PS1 data uploaded to Supernova Hunters between MJD 57570 and MJD 57586. This data set includes classifications from an initial beta test of the project prior to launch on MJD 57581.  A major contaminant is the presence of asteroids. These appear in the difference image as identical to supernovae, and are in that sense ``correctly classified'' but are identified here via cross-matching with the Minor Planet Center.

The results of the machine learning were additionally reviewed by at least one expert member of the team (normally DEW or KWS) to identify genuine supernovae.   Candidates were divided into `real' and `bogus' categories based on these expert classifications.  We note that a future improvement to the project would be to inject fake real and bogus detections into the data as a means to track performance.  The Andromeda Project \citep{Johnson15} and Space Warps are examples of two citizen science projects to have taken this approach, while \citet{Goldstein15} used faked detections to augement DES training data for their real-bogus classifier.  With this approach we would no longer be reliant on the assumption that every expert label is correct, but we must be careful to ensure that injected fake sources are truly representative of our observations.

\begin{figure}
   \includegraphics[width=84mm]{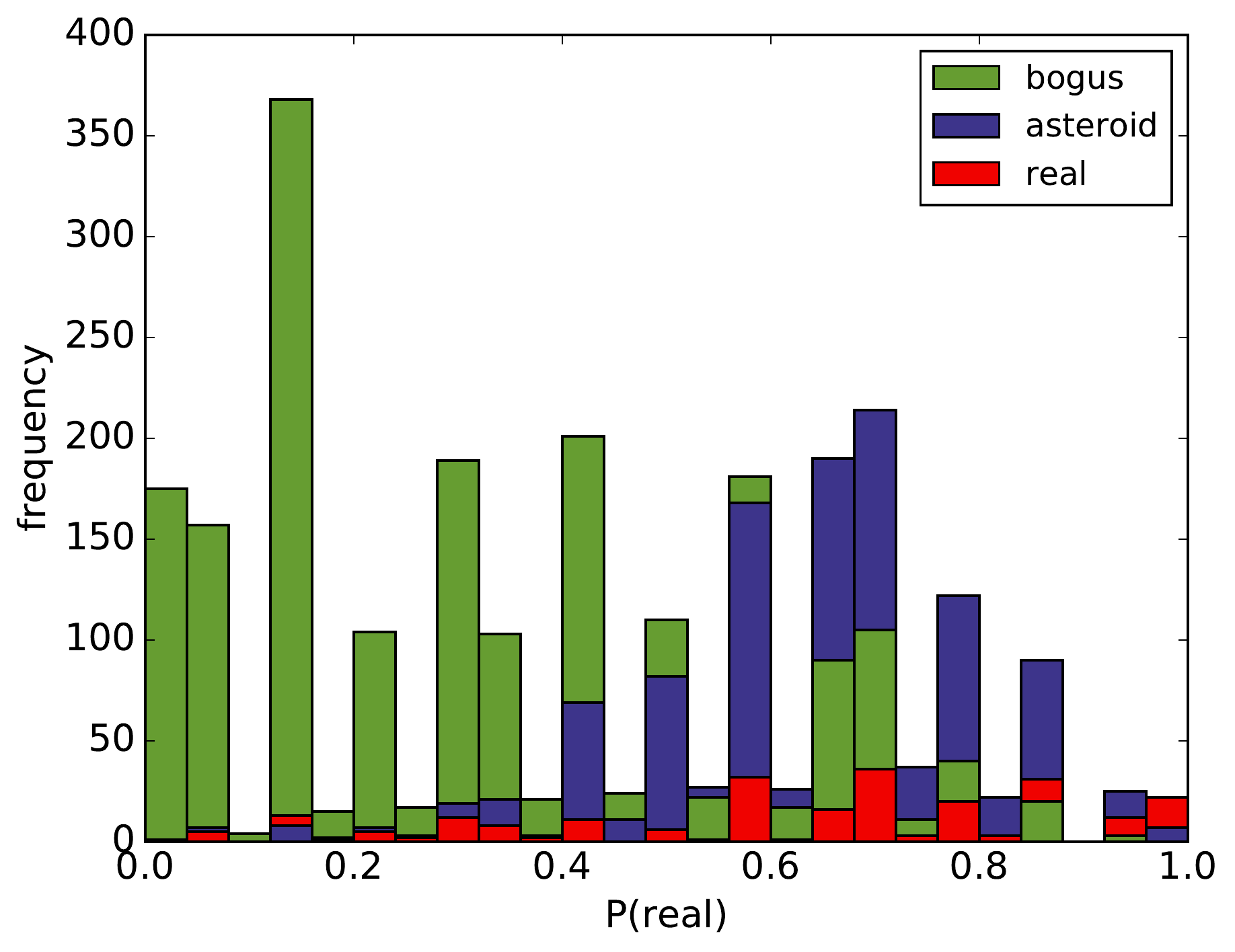}
   \caption{The distribution of $P(real)$ from Supernova Hunters for objects detected between 
            MJD 57570 and MJD 57586.  Compared with the machine $h(x)$ in Figure~\ref{fig:machine_dist}
            the objects at the extremes are pure.  There are no real detections with 
            $P(real) < 0.04$ and few bogus detections above 0.92.} 
   \label{fig:human_dist} 
\end{figure}

Candidates with high scores as assigned by the machine are more likely to be real. However, although the machine successfully rejects the majority of bogus candidates, the sample produced by the simple cut on hypothesis is far from pure; 1403 real candidates from 3384 in the sample. Higher cutoffs run the risk of rejecting an increasing number of real candidates; requiring a 1\% false positive rate will result in a missed detection rate of 60.3\%.  In order to improve this performance, candidates which exceed the $h(x) = 0.436$ threshold (see Section~\ref{sec:cnn}) were also classified by volunteers via the \emph{Supernova Hunters} project. The results of this analysis are shown in Figure~\ref{fig:human_dist}. 

Volunteer classifications were combined using the simplest possible metric; the fraction of volunteers who identified a detection as real is assumed to be an estimate of the probability of that candidate being real, denoted $P(real)$. Despite this simple procedure, the results show that volunteers could effectively distinguish between real and bogus classifications. However, the structure of the resulting distribution is strikingly different from that of the machine classifier. Whereas for machine classification, a threshold could be chosen to give a complete but not pure sample, with volunteer classifications it is easier to construct a pure sample of candidates which are highly likely to be supernovae, but this sample is far from complete. There are candidates judged `real' by experts even at low probabilities although there were no real candidates assigned $P(real) < 0.04$. 

There are two routes which might be expected to improve this performance.  First, we could improve on the naive combination of volunteer votes described above.  To this end citizen science projects typically explore methods to weight volunteer contributions (see, for example \citet{Schwamb12}, \citet{Willett13} and \citet{Marshall16}).  Second, given that we have a human and machine score for every detection we could seek a combination of the two in the hope of benefiting from the different capabilities of both.

\subsection{Combining human and machines} 
\label{sec:combo}
In a companion paper analysing performance of Galaxy Zoo, \citet{Beck17} used a machine classifier running in parallel to a simulation of the Galaxy Zoo 2 project \citep{Willett13}. They showed that the addition of such a classifier, which retired subjects classified above a certain level of confidence at the end of each day, retraining each time, can greatly accelerate the speed of classification in a data set. In their work, images are retired by either machine or human, whereas we set out in this section to use a combination. While they retrain their machine with volunteer input as it accumulates, we use a static training set derived from expert classifications.

\begin{figure}
   \includegraphics[width=84mm]{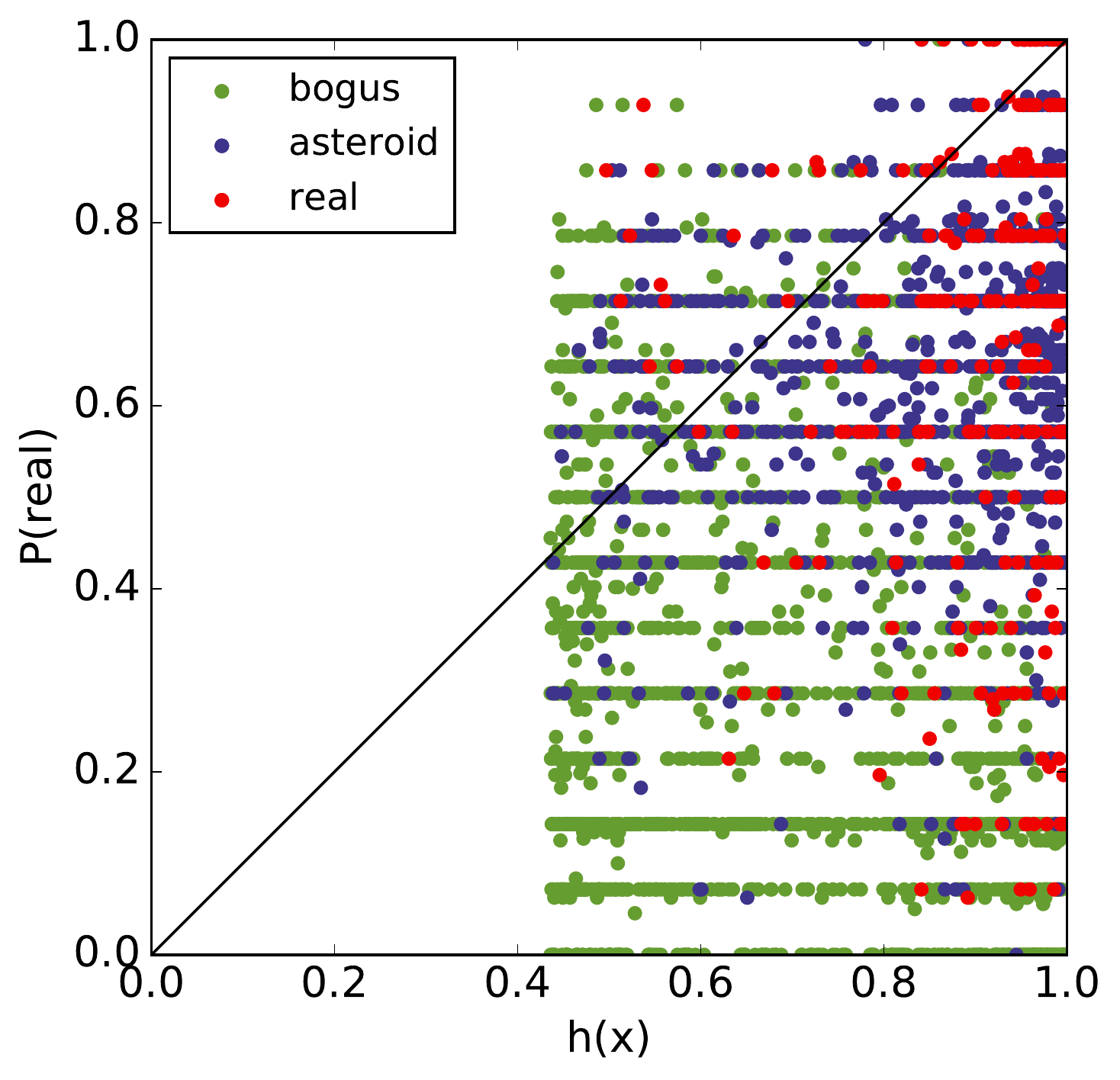}
   \caption{The $P(real)$ from Supernova Hunters against the machine $h(x)$ for 3384 detected 
            objects between MJD 57570 and MJD 57586.  $P(real)$ and $h(x)$ are combined by projecting the data onto 
             the solid black line in the euclidean sense.  A Spearman rank correlation test shows the correlation between $P(real)$ and $h(x)$ to be 0.237.}
   \label{fig:combo_train} 
\end{figure}

Figure~\ref{fig:combo_train} shows the combination of human and machine classifications. It is immediately apparent from the figure that no single threshold on either machine or human classification can outperform the combination of the two. This is an important result; it is the first time that the benefits of combining classification from both machines and volunteers has been clearly demonstrated using data from a live astronomical survey. 

How should the two independent classifications be combined?  We simply apply a decision boundary of the form $\tau = (x + y)/2$ on the 2D surface, where $0\leq\tau\leq1$.  For a constant value of $\tau$ a candidate is classified as bogus if $[h(x)+P(real)]/2 <= \tau$ and classified real otherwise.  This is equivalent to projecting the data onto $P(real)=h(x)$, producing a new scalar score for each detection and classifying candidates as bogus if the combined score is less than or equal to $\tau$.  

\begin{figure}
   \includegraphics[width=84mm]{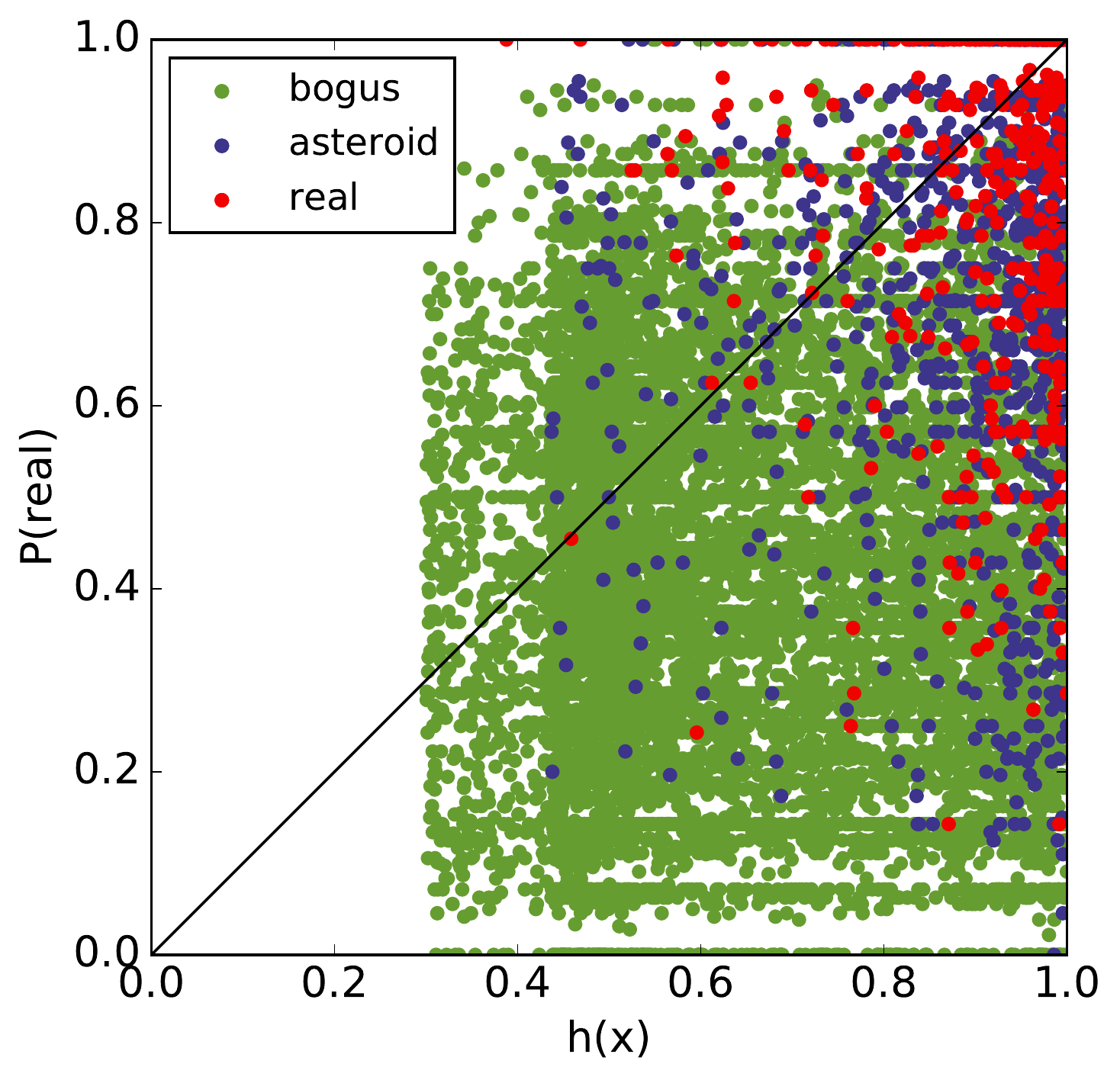}
   \caption{The same as Figure~\ref{fig:combo_train} but on a new sample of 10908 objects detected between
            MJD 57587 and MJD 57627.  For one week during this period we relaxed our cut on $h(x)$ to 0.3 which allowed us to recover a supernova with $h(x)$=0.39, but which achieved a $P(real)$=1.0 from Supernova Hunters.  In this case the Spearman rank correlation is found to be 0.122.}
   \label{fig:combo_test} 
\end{figure}

\begin{figure*}
   \begin{minipage}{160mm}
     \includegraphics[trim=35mm 35mm 35mm 35mm,clip,width=160mm]{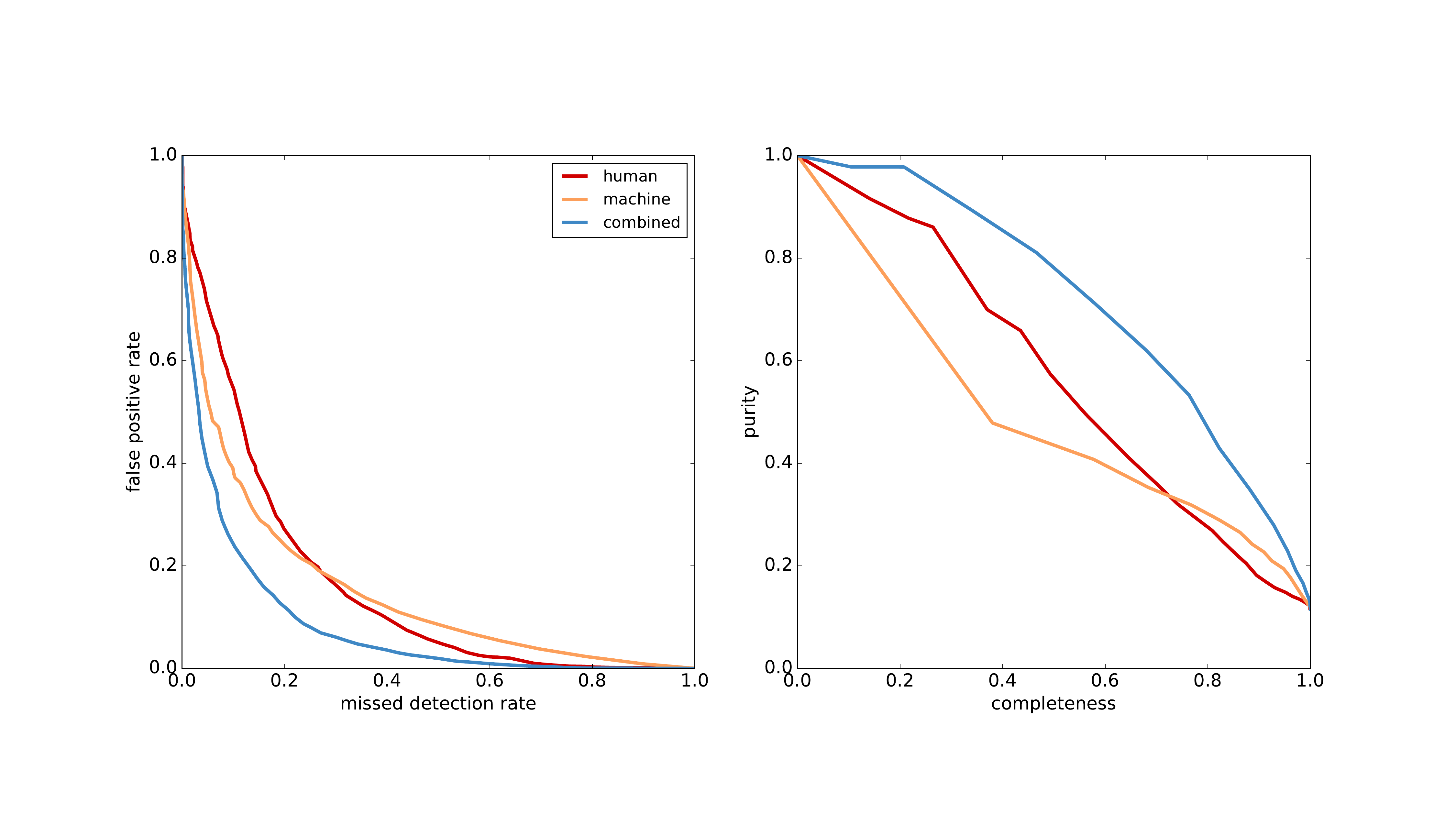}
   \caption{Left: ROC curve showing performance measured on data in Figure~\ref{fig:combo_test} for human (red), machine (yellow) and
            the combination of human and machine classifications (blue).  Right: The equivalent Purity-Completeness curve.  Both
            plots show that the combination always outperforms humans and the machine individually.} 
   \label{fig:roc} 
   \end{minipage}
\end{figure*}

As an independent test we apply this same method to data between MJD 57587 and MJD 57627 in
Figure~\ref{fig:combo_test}.  For Figure~\ref{fig:combo_train} and Figure~\ref{fig:combo_test} a Spearman rank correlation test gives 0.237 and 0.122 respectively, showing $P(real)$ and $h(x)$ to be only weakly correlated.  Between MJD 57609 and MJD 57615 we relaxed our cut on $h(x)$ from 0.436 to 0.3 uploading any objects passing this cut to Supernova Hunters.  This allowed us to explore the subjects wrongly classified by the machine and resulted in the recovery of SN 2016fev, a Type Ia supernova that would have been automatically rejected with $h(x)$=0.39, but which received a $P(real)$ of 1.0 from Supernova Hunters.  The performance of the combination method on this data set is shown in the Receiver Operator Characteristic (ROC) Curve and Purity-Completeness (Precision-Recall) Curves plotted in Figure~\ref{fig:roc}.

For any choice of false positive rate, the combination of classifications produced a lower missed detection rate. Equally, for any required purity or completeness the combination provides a better trade-off. 

We have chosen to implement one of the simplest methods for combining human and machine classifications to demonstrate how they complement one another, but it is easy to think of more complex combination methods. For example we trained a linear Support Vector Machine (SVM) on the data presented in Figure~\ref{fig:combo_train} and found, unsurprisingly, that the performance measured on the data in Figure~\ref{fig:combo_test} was typically within 1\% of the values reported in Tables~\ref{tab:roc_fpr} and ~\ref{tab:roc_mdr}.  Although the gains in this example are negligible, if we wish to incorporate additional information from an ensemble of machine classifiers for example, and the input space becomes higher dimensional, such methods become important.  It is unlikely that the combination method presented here will work for higher dimensional data, but we can expect that an SVM may take advantage of the additional information.

\subsection{Improving $P(real)$ with SWAP} 
We also expect to gain from improving $P(real)$.  To demonstrate this we implemented the Space Warps Analysis Pipeline (SWAP) \citep{Marshall16}, an algorithm designed to improve the sample of good gravitational lens candidates from the Space Warps\footnote{\url{https://spacewarps.org}} citizen science project in images from the Canada---France---Hawaii Telescope Legacy Survey (CFHTLS) \citep{Gwyn12}.  SWAP does not treat all classifiers equally, but quantifies the value of their contribution in terms of the information gained.  SWAP assigns a software \emph{agent} to individual citizen scientists.  Classifications are weighted according to the agent's estimate of likely performance based on each volunteer's past performance measured on gold standard data.  The agent maintains a confusion matrix that monitors the fraction of gold standard data correctly or incorrectly classified by a volunteer for each class (``LENS'' or ``NOT'' in the case of Space Warps).   We use the online implementation of SWAP considering each classification in turn.  We make several small adjustments to the original SWAP  implementation.  First, our gold standard labels are ``real'' and ``bogus'' rather than ``LENS'' or ``NOT''.  Second, we set the prior probability that each detection is a real supernova $\rho_0$ to be 0.01, roughly the expected ratio between real and bogus detections each night determined by expert classifications.  Finally, in our analysis we do not set rejection and detection thresholds, instead we continue to require that at least seven volunteers classify each subject.  As in \citet{Marshall16} we set the initial confusion matrix for the $i-$th volunteer to:\\

\begin{figure*}
   \begin{minipage}{160mm}
     \includegraphics[trim=55mm 45mm 55mm 55mm,clip,width=160mm]{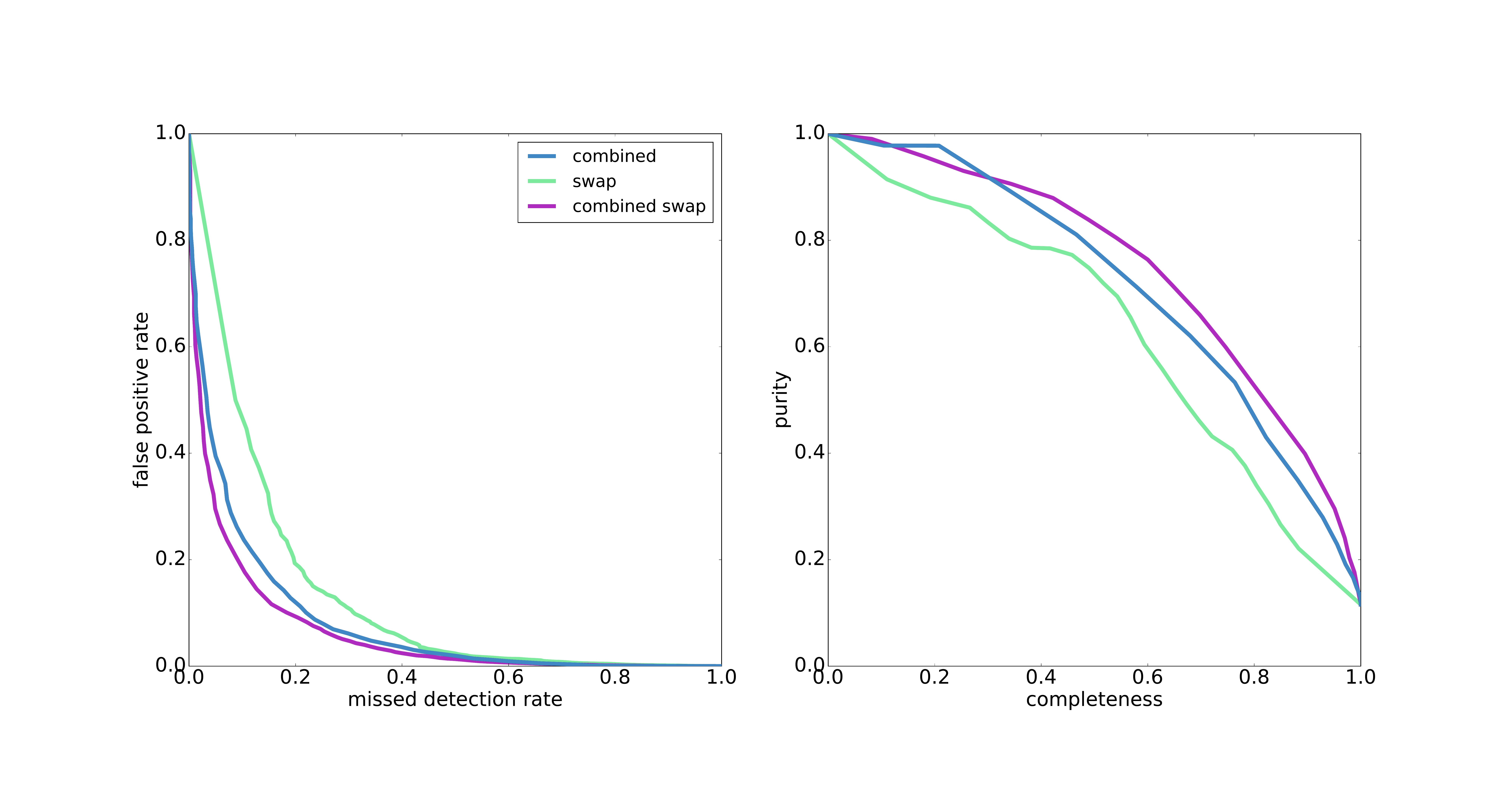}
   \caption{Left: ROC curve for different approaches to classification.  Right: The corresponding Purity-Completeness curves.  The blue lines are the same as in Figure~\ref{fig:roc} showing the simple combination of the machine with the basic human score described in Section~\ref{sec:perform}.  The green line shows the results of implementing SWAP, which improves on the basic human score (red line in Figure~\ref{fig:roc}).  Combining this improved score with the machine produces the pink line.}
   \label{fig:roc_1} 
   \end{minipage}
\end{figure*}

$\bm{M}^i = \begin{bmatrix}
                        0.5&0.5\\
                        0.5&0.5 \\
                        \end{bmatrix}$.\\
\\
This initialisation of the confusion matrix corresponds to that of a random classifier, but will be quickly modified as we observe classifications of gold standard data by this volunteer.  The results of applying SWAP to the citizen scientist classifications on data between MJD 57587 and MJD 57627 (Figure~\ref{fig:combo_test}) are shown as the purple line in Figure~\ref{fig:roc_1}.  Compared with the human performance in Figure~\ref{fig:roc}, SWAP considerably improves $P(real)$.  By combining this improved score from human classifications with the machine, using the same method as above, we once again observe an improvement in the measured performance (pink line from Figure~\ref{fig:roc}).  In fact, apart from a small exception at about 97\% purity and 25\% completeness (likely an artefact of the specific data set) the combination outperforms all previous methods for classifying this data set, demonstrating that an improvement in either $P(real)$ or $h(x)$ can lead to valuable performance gains when combined.  SWAP combined with the machine produces a few percent improvement over all the other methods tested so far for any figure of merit in Tables~\ref{tab:roc_fpr} and ~\ref{tab:roc_mdr}.

\begin{figure}
   \includegraphics[trim=0mm 0mm 0mm 0mm,clip,width=84mm]{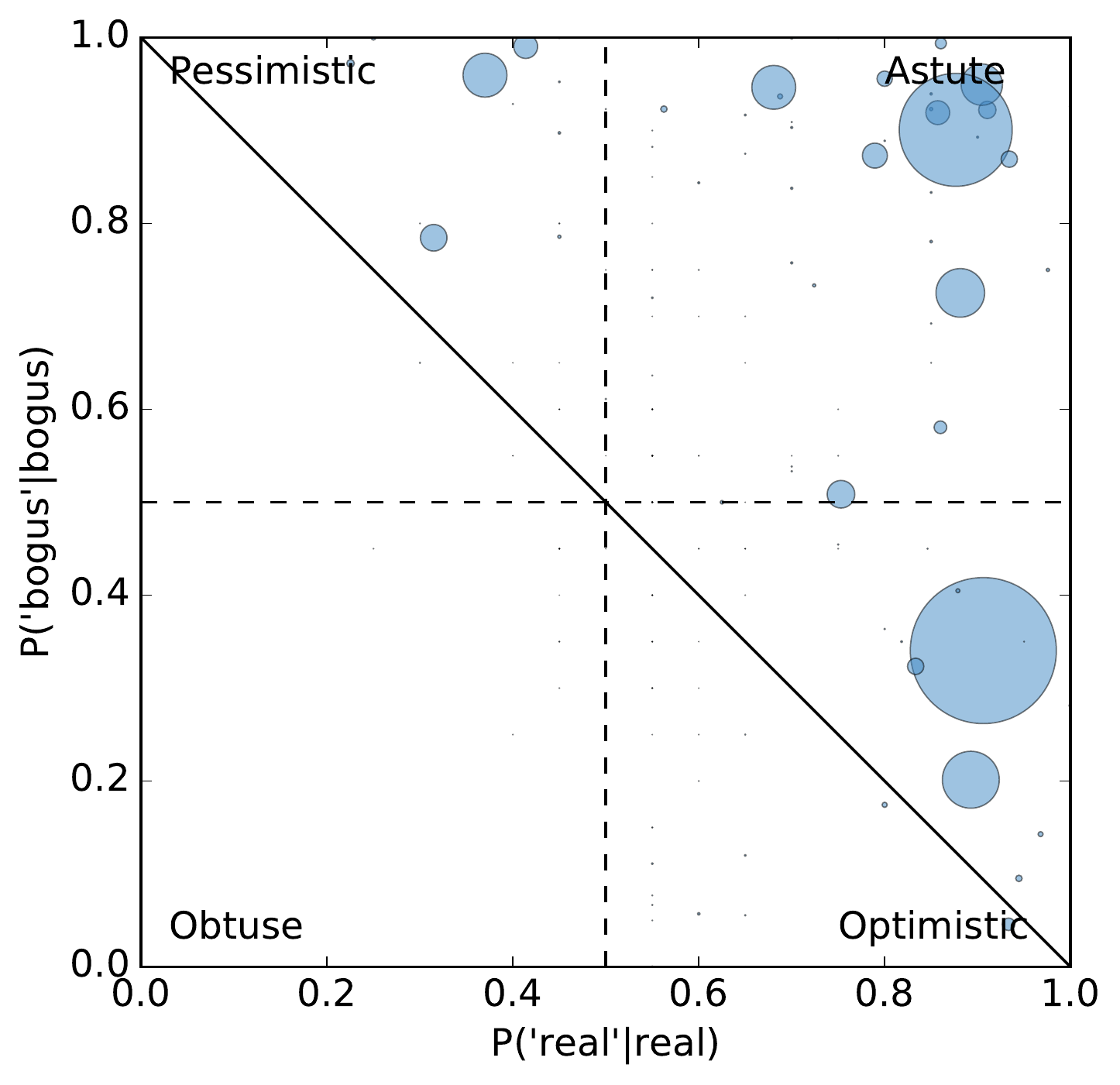}
   \caption{Confusion matrix elements for volunteers classifying detections from Figure~\ref{fig:combo_test} with apparent magnitude $13\leq m < 18$.  Each point represents an individual volunteer, where the size of the point corresponds to the quantity of gold standard data in this magnitude bin classified by the volunteer.  Larger points therefore correspond to more experienced classifiers.  Many volunteers tend to lie high in the top right of the ``Astute'' region, a perfect classifier would lie at (1,1).  This shows that citizen scientists can accurately classify high signal-to-noise real and bogus detections.}
   \label{fig:13-18} 
\end{figure}

\begin{figure}
   \includegraphics[trim=0mm 0mm 0mm 0mm,clip,width=84mm]{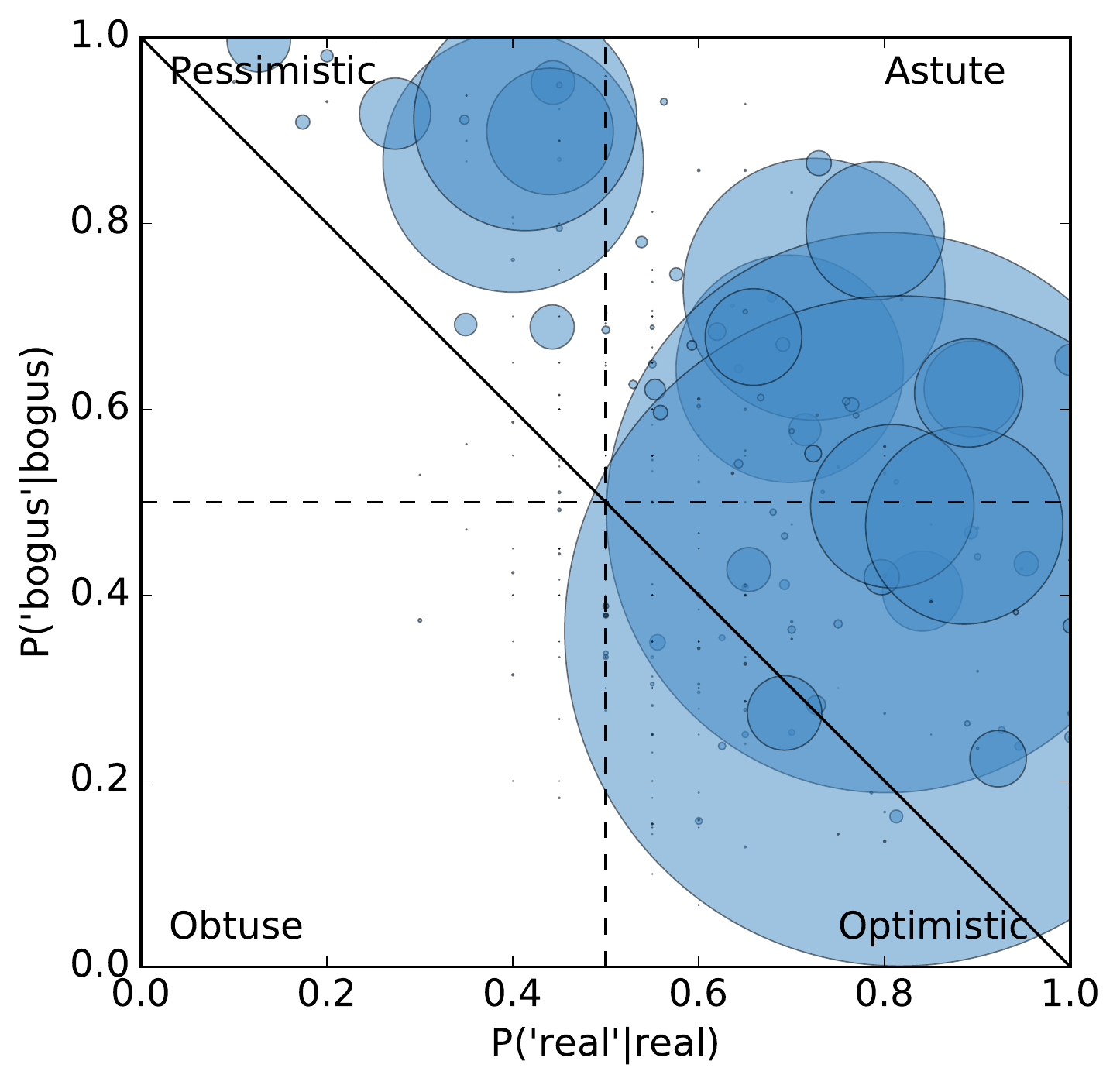}
   \caption{Similar to Figure~\ref{fig:13-18} but for detections with apparent magnitude $19\leq m < 20$.  In this case volunteers are further from the top right and tend towards ``Pessimistic'' or ``Optimistic'' classifiers.}
   \label{fig:20} 
\end{figure}

Considering the scenario where we are reliant on citizen scientists to label training data for machine learning algorithms, it seems prudent to consider how we might further improve $P(real)$ in the future.  One observation we make is that citizen scientists perform differently on detections with varying degrees of signal-to-noise.  In general, citizen scientists are very good at recovering bright supernovae but find it more and more difficult as detections get closer to the detection limit.  Given the history of classifications of gold standard data a volunteer has submitted, we can calculate the probability the volunteer will classify a detection as real given that the true label is real, $P(``real"|real)$, by simply taking the fraction of the gold standard detections labelled ``real'' the volunteer correctly classified. We can similarly calculate $P(``bogus"|bogus)$.  These are the probabilities tracked by the SWAP agent confusion matrix and which we have plotted on the unit plane for the 3158 volunteers participating in the project at this time for different magnitude bins in Figure~\ref{fig:13-18} and Figure~\ref{fig:20}.  A volunteer who correctly classifies every gold standard example they have seen is a perfect classifier and will lie at (1,1) in the ``Astute'' quadrant of the plot.  The size of the point corresponds to the quantity of gold standard data classified by that volunteer in a given magnitude bin.  Most volunteers have submitted only a few classifications and cannot be seen on the plot.  Clearly volunteers are more ``Astute'' at classifying brighter sources, tending to lie closer to the top right of the plot. This information is lost in the basic SWAP implementation.  A volunteer's classification could now be weighted according to the magnitude bin that a given detection falls in.  This results in a simple modification of the SWAP calculation of $P(real)$.\\
\\
In this case we are taking advantage of additional ``metadata" that is available for each detection and assuming that it has some effect on classifiers' behaviour.  We also expect that humans and machines are good at classifying different types of images.  In our case the CNN is good at classifying detections around 20th magnitude, but often misclassifies detections brighter than 17th magnitude.  This is because the training set is dominated by fainter detections.  The machine has not been able to learn a function that accurately maps between the pixels and the classification for bright examples because it has not ``seen'' many during training.  In Tables~\ref{tab:roc_fpr} and ~\ref{tab:roc_mdr} we include the results of running SWAP on the magnitude spilts and the combination with the machine scores.  We find that, for the most part, this gives a $\sim$1\% improvement in performance over not using metadata splits depending on the Figure of Merit.  There are other metadata parameters we could explore, both properties of the survey (seeing) and properties of the detection (proximity to a galaxy).  Tables~\ref{tab:roc_fpr} and ~\ref{tab:roc_mdr} also show the effect of using seeing for the metadata splits and we observe similar performance to magnitude.  Although the gains are small this may indicate that a more thorough analysis of metadata is worth pursuing.  For instance we could more carefully define where the splits on metadata should be made and consider the effect of combining multiple metadata parameters.  In future, it would be interesting to train an ensemble of machines with each only consuming retired data from a single metadata split.  When adequately trained, the resulting machines would specialise in specific regions of parameter space and could eventually be folded back into the project.  We could also incorporate these machines into the SWAP analysis with their classifications weighted just as for volunteers.

\begin{table*}
\begin{minipage}{150mm}
\centering
\begin{tabular}{|c|c|c|c|c|c|c|c|c|c|}
       &             &               &                 &           & SWAP   & SWAP     & combined  & combined     & combined      \\
FPR & human & machine & combined & SWAP &  (mag.) & (seeing) &SWAP          & SWAP (mag.) & SWAP (seeing)\\
\hline
1\% & 73.9\% & 90.1\% & 58.7\% & 66.6\% & 64.5\% & 64.2\% & 54.5\% & 54.1\% & 53.5\% \\
5\% & 56.3\% & 69.7\% & 35.8\% & 41.0\% & 39.7\% & 40.3\% & 30.2\% & 29.7\% & 28.8\%\\
10\% & 45.6\% & 46.7\% & 23.8\% & 31.2\% & 30.0\% & 30.5\% & 20.5\% & 18.9\% & 20.1\%\\
\end{tabular}
\caption{Missed detection rate recorded for a choice of false positive rates, based on expert classifications.}
\label{tab:roc_fpr}
\end{minipage}
\end{table*}
%
\begin{table*}
\begin{minipage}{150mm}
\centering
\begin{tabular}{|c|c|c|c|c|c|c|c|c|c|}
       &             &               &                 &           & SWAP   & SWAP     & combined  & combined     & combined      \\
MDR & human & machine & combined & SWAP &  (mag.) & (seeing) &SWAP          & SWAP (mag.) & SWAP (seeing)\\
\hline
1\% & 92.5\% & 85.9\% & 69.3\% & 100.0\% & 100.0\% & 100.0\% & 66.2\% & 68.7\% & 71.2\%\\
5\% & 75.1\% & 52.8\% & 41.8\% & 100.0\% & 100.0\% & 100.0\% & 29.6\% & 31.0\% & 31.7\%\\
10\% & 53.8\% & 39.1\% & 26.5\% & 49.9\% &  47.9\% & 49.1\% & 20.6\% & 19.6\% & 20.2\%\\
\end{tabular}
\caption{False positive rate recorded for a selection of missed detection rates, based on expert classifications.
}\label{tab:roc_mdr}
\end{minipage}
\end{table*}

\section{Conclusions}
\label{sec:conclusions}

In this paper we introduced a new citizen science project, Supernova Hunters built entirely with the off-the-shelf Zooniverse Project Builder\footnote{\url{https://www.zooniverse.org/lab}} requiring no custom features or additional development.  The project aims to classify detections of potential supernova candidates from the Pan-STARRS Survey for Transients either as real transients or bogus detections of image processing or instrumentation artefacts.  To be uploaded to the Supernova Hunters project a detection must first pass a series of cuts based on catalogue information and secondly be promoted by our machine learning algorithm, a Convolutional Neural Network. With this approach we expect that only about 5\% of the false positives passing the catalogue cuts make it into the project, greatly reducing the number of objects we ask volunteers to screen.  Citizen scientists excel at mining this data for a very pure sample of high $P(real)$ supernova candidates, typically those of higher signal-to-noise and offset from a galaxy.  But compared with expert labels many less obvious candidates are missed.  Rather than immediately consider methods of weighting classifications of individual citizen scientists, we instead applied a simple combination of the scores provided by humans and machines.  We showed the new combined score achieved better performance than either individually for any choice of purity or completeness.

We expect that there are many ways to improve.  As an example we generalised the Space Warps Analysis Pipeline (SWAP) to Supernova Hunters and applied it to the volunteer classifications and again combined the resulting scores with those from the machine.  We found that this resulted in a few percent performance gain compared to the next best method.  To further improve $P(real)$ we could invest more effort into educating citizen scientists to identify more subtle artefact indicators.  For example the Gravity Spy\footnote{\url{https://www.zooniverse.org/projects/zooniverse/gravity-spy}} project has implemented a training regimen where volunteers are provided with feedback on classifications of gold standard data and can progress to more advanced levels performing more complex tasks \citep{Zevin17}.  This could help address the many bogus detections with $P(real)>0.5$, but relatively few real detections below 0.5 in Figure~\ref{fig:human_dist}. Another improvement could be with our machine classifier, which was trained at the beginning of the PSST survey and the algorithm was specifically chosen to learn from the limited amount of training data available.  Given the large volume of data accumulated since, we could train more sophisticated algorithms that can learn more complex relationships between the features, though extracting robust labels for this additional data is a challenge that still needs to be addressed. 

This effort offers hope for dealing with the large data volumes from all sky surveys such as LSST, ZTF \citep{Bellm14}, ATLAS \citep{Tonry16} and Pan-STARRS2.  We used machines to reject the vast majority of false positives and then combined the machine hypotheses with classifications from a few thousand citizen scientists for the remaining candidates.  With Supernova Hunters we have not actively sought additional citizen scientists, beyond the $\sim30000$ volunteers on the Zooniverse beta testing e-mail list, who were asked to review the project before launch.  New volunteers must ``discover'' the project on the Zooniverse projects page to participate.  Given that we could actively seek the participation of $\sim 10^6$ registered Zooniverse volunteers and assuming that 10\% chose to participate, with the current classification rate (21000 in the first 24 hours each week from $\sim$6000 volunteers) we could achieve $\sim 350000$ classifications per night.  This provides 0.35 classifications for the $\sim 10^6$ transient alerts expected from LSST at the beginning of the survey \citep{Ridgway14}.  If the false positive rate is an order of magnitude more than the transient alert rate (perhaps overly pessimistic given the expected $\sim$ 500-2200 false positives per field per visit \citep{Becker13} with a $5\sigma$ detection threshold) and assuming we can discard 90\% of those with machine learning we can expect to achieve 0.175 citizen science classifications per promoted detection.  Assuming we will require roughly 10 classifications per detection before considering it classified, we are roughly two orders of magnitude short.  Making up this deficit may be achievable with continued improvements to difference imaging \citep{Zackay16}, automated real-bogus classification, encouraging greater participation from a growing community of citizen scientists and more efficient use of their classifications.

\section*{Acknowledgments}
The Pan-STARRS1 Surveys have been made possible through contributions of the Institute for Astronomy, the University of Hawaii, the Pan-STARRS Project Office, the Max-Planck Society and its participating institutes, the Max Planck Institute for Astronomy, Heidelberg and the Max Planck Institute for Extraterrestrial Physics, Garching, The Johns Hopkins University, Durham University, the University of Edinburgh, Queen's University Belfast, the Harvard-Smithsonian Center for Astrophysics, the Las Cumbres Observatory Global Telescope Network Incorporated, the National Central University of Taiwan, the Space Telescope Science Institute, the National Aeronautics and Space Administration under Grant No. NNX08AR22G issued through the Planetary Science Division of the NASA Science Mission Directorate, the National Science Foundation under Grant No. AST-1238877, the University of Maryland, and Eotvos Lorand University (ELTE) and the Los Alamos National Laboratory.  The development of the Zooniverse platform was supported by a Global Impact Award from Google.  We also acknowledge support from STFC under grant ST$/$N003179$/$1.  LF and MB were partially supported by the National Science Foundation under Grant No. AST-1413610.  DW and LF were partially supported by the National Science Foundation under Grant No. IIS-1619177.  We also wish to acknowledge the dedicated effort of our citizen scientists who have made this work possible.

\bibliographystyle{mnras} 
\bibliography{ps1_real_bogus}

\bsp	
\label{lastpage}

\end{document}